\documentclass[journal=jpcbfk,manuscript=article,layout=twocolumn]{achemso}

\usepackage[version=3]{mhchem} 
\usepackage{gensymb}
\usepackage{siunitx}
\usepackage{amsmath}
\usepackage{enumerate}
\usepackage{subfigure}
\usepackage{dblfloatfix}

\newcommand{\bq}{\begin{equation}} \newcommand{\eq}{\end{equation}}
\newcommand\Int{\operatorname{int}}
\newcommand\gas{\operatorname{gas}}
\newcommand\sheet{\operatorname{sheet}}
\newcommand\barrier{\operatorname{barrier}}
\newcommand\TS{\operatorname{TS}}
\newcommand\IIS{\operatorname{SS}}
\newcommand\ad{\operatorname{ad}}

\newcommand\x{\operatorname{x}}

\newcommand\D{\operatorname{d}}
\newcommand\T{\operatorname{T}}
\newcommand\coh{\operatorname{coh}}

\renewcommand{\thefootnote}

\author{Parham Rezaee}
\email{parham.rezaee@alum.sharif.edu}
\affiliation[Sharif University of Technology]
{Department of Chemistry, Sharif University of Technology, Tehran, Iran}
\author{Hamid Reza Naeij$^{\dagger}$}
\affiliation[Sharif University of Technology]
{Department of Chemistry, Sharif University of Technology, Tehran, Iran}
\title
  {Graphenylene--1 Membrane: An Excellent Candidate for Hydrogen Purification and Helium Separation}

\let\oldmaketitle\maketitle
\let\maketitle\relax
\begin{document}

\twocolumn[
\begin{@twocolumnfalse}
\oldmaketitle
\begin{abstract}
In this study, we use the density functional theory (DFT) calculations and the molecular dynamics (MD) simulations to investigate the performance of graphenylene--1 membrane for hydrogen (\ce{H2}) purification and helium (\ce{He}) separation. The stability of this membrane is confirmed by calculating its cohesive energy. Our results show that a surmountable energy barrier for \ce{H2} (\SI{0.384}{\electronvolt}) and \ce{He} (\SI{0.178}{\electronvolt}) molecules passing through graphenylene--1 membrane. At room temperature, the selectivity of \ce{H2}/\ce{CO2}, \ce{H2}/\ce{N2}, \ce{H2}/\ce{CO} and \ce{H2}/\ce{CH4} are obtained as \num{3d27}, \num{2d18}, \num{1d17} and \num{6d46}, respectively. Furthermore, we demonstrate that graphenylene--1 membrane exhibits the permeance of \ce{H2} and \ce{He} molecules are much higher than the value of them in the current industrial applications specially at temperatures above 300 \si{\kelvin} and 150 \si{\kelvin}, respectively. We further performed MD simulations to confirm the results of DFT calculations. All these results show that graphenylene--1 monolayer membrane is an excellent candidate for \ce{H2} purification and \ce{He} separation. 
\end{abstract}
\end{@twocolumnfalse}
]



{\fnsymbol{footnote}}
\footnotetext[1]{$\dagger$E-mail: naeij@alum.sharif.edu}
\section{Introduction}

Nowadays, depletion of the fossil fuel and increased environmental pollution have been a global concern. It seems that it is essential to exploit renewable and clean energy instead of the fossil fuels. Recently, \ce{H2} is regarded as one of the most efficient substitutes of the fossil fuels because of its high energy content, clean burning product, natural abundance and  renewable nature. Therefore, it will be the most attractive and promising energy source in the future \cite{Winter, Andrews, Schlapbach,Tollefson,Turner,Park}.

Currently, steam-methane reforming (\ce{CH4 + H2O -> CO (CO2) + H2}) is the common technology for \ce{H2} production in industry, which is an economical process at a large scale \cite{Pen, Zhu}. So, there are impurities of \ce{CO}, \ce{CH4}, \ce{CO2} and \ce{N2} in this reaction. These impurities can severely cause bad influences on energy content, storage and utilization of \ce{H2} \cite{Freemantle, Oetjen, Alves}. Consequently, high-quality purification of \ce{H2} is very important and finding an effective and low-cost approach for it is very challengable.

Moreover, \ce{He} is an irreplaceable natural resource. Although, it is the second most abundant element in the universe, the shortage of it will become increasingly serious \cite{Li 1, Zhu 1, Gao}. Today, demand of \ce{He} in various industrial and scientific applications such as cryogenic science and silicon-wafer manufacturing is increasing \cite{Nuttall, Cho}. In addition, the isotope, \ce{^{3}He}, is important for fundamental researches in physics \cite{Halperin}. It seems that development of low-cost energy strategy for \ce{He} separation from natural gases is highly desired. 
 
The common traditional \ce{H2} purification technologies are cryogenic distillation and pressure swing adsorption. However, these technologies have disadvantages such as complicated operation and high energy consumption. Recently, membrane separation technology has been widely used because of low energy consumption, low investment cost and facile operation \cite{Jiao, David,Deng, Bernardo,Spillman}. In this regrad, many advanced membrane materials, ranging from tens of nanometres to several micrometres in thickness, have been developed for gas separation, including polymeric membranes \cite{Lin}, metal organic framework \cite{Herm}, dense metal membranes \cite{Chandrasekhar}, zeolite membranes \cite{Li 2} and etc.
 
Generally, a suitable \ce{H2} purification membrane should have two important characteristics: (1) at least interaction between \ce{H2} and membrane; (2) a certain energy barrier to distinguish between \ce{H2} and other gases. Therefore, the selectivity and permeance are two important factors to evaluate the performance of \ce{H2} purification membrane \cite{Ji}. An ideal two-dimensional (2D) membrane for gas separation would show an appropriate balance between the selectivity and the permeance factors. However, traditional membrane usually encounter the selectivity-permeance trade-off problem \cite{Robeson 1, Robeson 2, Gao 2}. The permeance is inversely proportional to the thickness of the membrane. So, one-atom thin membrane can be an ideal candidate for \ce{H2} purification \cite{Chang}.

The design and synthesis of suitable 2D membranes for gas separation have attracted wide attention in theoretical and experimental contexts \cite{Kang, Giraudet,Jiao 1}. Recently, graphene and carbon allotropes related sp–sp$^2$ hybridization have been proposed as the membranes for gas separation \cite{Schrier,Liu}. These structures are formed by benzene rings linked with acetylenic linkages, graphyne, or diacetylenic linkages, graphdiyne. In an important study, Liu {\it et al.} synthesized graphenylene--1 and graphenylene--2 by replacing ethynylene groups of $\gamma$-graphyne and graphdiyne with $p$-phenylene units, respectively \cite{Liu2}. These new structures have  purely sp$^2$-hybridized 2D carbon structures with a hexagonal symmetry.

Many chracteristics of these carbon allotropes monolayer membranes such as highly diversified structures, periodically distributed uniform pores and high chemical and mechanical stability make them an appropriate candidate for the gas separation process \cite{Jiao, Bartolomei}. It is noteworthy that the acetylenic linkages in the structure of graphyne creates periodic pores, which the size of them can be controlled by the length of the linkages \cite{Kang 1}.

Recently, many studies have been done to investigate the gas separation process through carbon allotropes monolayer membarnes. For example, Cranford {\it et al.} performed MD simulations to obtain the temperature and pressure dependence of the gas purification in graphdiyne \cite{Cranford}. Zhang {\it et al.} showed that graphyne--1 is unsuitable membrane for \ce{H2} purification due to small pore size and insuperable diffusion energy barrier. In addition, graphdiyne, with larger pores, demonstrates a high selectivity for \ce{H2} over large gas molecules such as \ce{CH4}, but a relatively low selectivity over small molecules such as \ce{CO} and \ce{N2} \cite{Zhang}. To conquer this problem, Li {\it et al.} demonstrated that Ca-decorated graphyne has good \ce{H2} storage capacity \cite{Li 3}. Moreover, Sang {\it et al.} designed two dumbbell-shaped graphyne membrane and used DFT calculations and MD simulations to invetigate the performance of it. Their calculations showed that the designed membrane is suitable for \ce{H2} separation \cite{Sang}.

The question that arises here is that: Can we further enhance the performance of graphyne-based membrane for \ce{H2} purification? It seems that the pore size of graphyne is an important factor. So, we can obtain appropriate pore sizes by changing the elementary structure of graphyne to improve the selectivity and the permeance factors for \ce{H2} purification process. 

In the present study, we use DFT calculations and MD simulations to investigate the performance of graphenylene--1 monolayer membrane which synthesized by Liu {\it et al.} \cite{Liu2}, for \ce{H2} purification and \ce{He} separation. First, we examine the stability of this monolayer membrane by calculating its cohesive energy. Then, the energy barriers of the gas molecules passing through graphenylene--1 membrane were calculated using DFT calculations to obtain the selectivity and the permeance of the membrane. Moreover, the results of DFT calculations were confirmed by MD simulations. Our results show that high selectivity and excellent permeance for \ce{H2}/gas (\ce{CO}, \ce{CH4}, \ce{CO2}, \ce{N2}) and \ce{He}/gas (\ce{H2}, \ce{CO}, \ce{CH4}, \ce{CO2}, \ce{N2}, \ce{Ne}, \ce{Ar}) at differrent temperatures. 

\section{Computational Methods}

DFT calculations were carried out to optimize the structure of graphenylene--1 monolayer, compute its stability, calculate the energy barrier of the gas molecules passing through it and describe the electron density iso-surfaces for the gases interacting with porous graphenylene--1 monolayer. The Perdew-Burke-Ernzerhof (PBE) function under the generalized gradient approximation (GGA) is employed by the spin-unrestricted all-electron DFT calculations which interprets the nonhomogeneity of the true electron density using the gradient of charge density for exchange-correlation function. We adopt a dispersion correction for DFT calculations with Grimme’s method by adding a semi-empirical dispersion potential. The double numerical plus polarization (DNP) basis set is used to expand electronic wave function. The self-consistent field (SCF) calculations are performed with a convergence criterion of \num{1d-6} a.u. on the total energy to ensure the high-quality  results. In  addition, a  real-space  global orbital cutoff radius of \SI{4.5}{\angstrom} and a smearing point of 0.002 Ha are chosen in all calculations. The Brillouin zone is expressed using a \num{6 x 6 x 1} Monkhorst-Pack meshes. A \SI{20}{\angstrom} vacuum thickness is used to prevent the interaction between two sheets. A large 2D sheet \SI[product-units = power]{20.15 x 20.15}{\angstrom}{ in xy plane including 244 atoms of C and H is constructed to represent the 2D graphenylene--1 atomic layer. Iso--electron density surfaces were obtained by the Gaussian 09 program \cite{Frisch} at the B3LYP/6-31G(d) level with the D3 correction \cite{Tian} were plotted at isovalues \SI{0.007}{\elementarycharge\angstrom^{-3}} to determine the pore size of the membrane. On the basis of this method, we find the potential energy curves of a single \ce{H2}, \ce{He}, \ce{Ne}, \ce{CO2}, \ce{CO}, \ce{N2}, \ce{Ar} and \ce{CH4} molecule passing through the pore center of the membrane. 

Moreover, MD simulations were performed to analyze \ce{H2} purification and \ce{He} separation using Forcite code in Material Studio software under a canonical (NVT) ensemble conditions and the temperature (in the range of 200-600 \si{\kelvin}) was controlled by the Anderson thermostat. Periodic boundary conditions were applied in all three dimensions. The interatomic interactions were explained by a condensed-phase optimized molecular potential for atomistic simulation studies (COMPASS) force field \cite{Sun}, which has been widely used to compute the interactions between the gases and the carbon-based membranes \cite{Shan,Wu,Xu}. The cutoff distance of the Van der Waals interactions was set as \SI{12.5}{\angstrom} and the particle-particle particle-mesh (PPPM) technique used to calculate the electrostatic interactions. The cubic boxes with the dimensions of $59.0 \times 49.6 \times 200$ \si{\angstrom^3} for simulations were trisected along the z-direction with two pieces of graphenylene--1 membrane, forming one gas reservoir in the middle and two vacuum regions on both sides. During MD simulations, the carbon atoms on the edge of membrane were always fixed, while all the other atoms in the system were fully relaxed. The gas mixtures involved 120 \ce{H2} molecules, 40 \ce{H2O} molecules, 40 \ce{CO2} molecules, 40 \ce{N2} molecules, 40 \ce{CO} molecules and 40 \ce{CH4} molecules. The total simulation time was \SI{1000}{\pico\second}, and Newton’s equations were integrated using \SI{1}{\femto\second} timesteps.

\section{Results and Discussion}

\begin{table*}[!b]
\small
\centering
\caption{Kinetic diameter, adsorption height, adsorption energy and energy barrier of the gas molecules passing through graphenylene--1 membrane.}
 \label{tbl:kindi}
  \begin{tabular*}{\textwidth}{@{\extracolsep{\fill}}lcccccccc}
    \hline
      &  \ce{He}  & \ce{Ne} & \ce{H2} & \ce{CO2} & \ce{Ar} & \ce{N2} & \ce{CO} & \ce{CH4}\\
    \hline
     ${D_0}$(\si{\angstrom})\cite{B802426J} & 2.60 &  2.75  & 2.89 & 3.30 & 3.40 & 3.64 & 3.76 & 3.80 \\
     ${H_{\ad}}$(\si{\angstrom}) & 3.175 & 3.275 & 3.300 & 4.550 & 3.725 & 3.850 & 3.875 & 4.550 \\
     ${E_{\ad}}$(\si{\electronvolt}) & -0.027 & -0.074 & -0.077 & -0.114 & -0.118 & -0.113 & -0.119 & -0.114 \\
     ${E_{\barrier}}$(\si{\electronvolt}) & 0.178 & 0.296 & 0.384 & 2.017 & 2.271 & 1.472 & 1.407 & 3.167\\
    \hline
  \end{tabular*}
\end{table*}

The optimized structure of a 2D graphenylene--1 monolayer membrane is shown in FIG. \ref{fcrys}. The optimized crystal lattices are $a$ = $b$ = \SI{8.49}{\angstrom} and actually, graphenylene--1 monolayer membrane can be seen as benzene rings linked together, resulting in a 2D network with periodically distributed pores. In graphenylene--1 monolayer, the length of C--C bonds involved in vertical benzene rings is \SI{1.40}{\angstrom} while the length of C--C bonds presented in in-plane benzene rings is \SI{1.39}{\angstrom}. All linked C--C bonds have a uniform length of \SI{1.46}{\angstrom} which are in good agreement with the previous theoretical predictions\cite{shekar_interlocked_2017,owais_selective_2018,treier_surface-assisted_2011}. Moreover, we also show the electron density iso--surfaces of pores in graphenylene--1 monolayer membrane. As shown in FIG. \ref{fcrys}, the pore of this monolayer membrane have a trigonal shape with the size of \SI{2.20}{\angstrom}. 

\begin{figure}[t]
\centering
\includegraphics[scale=0.27]{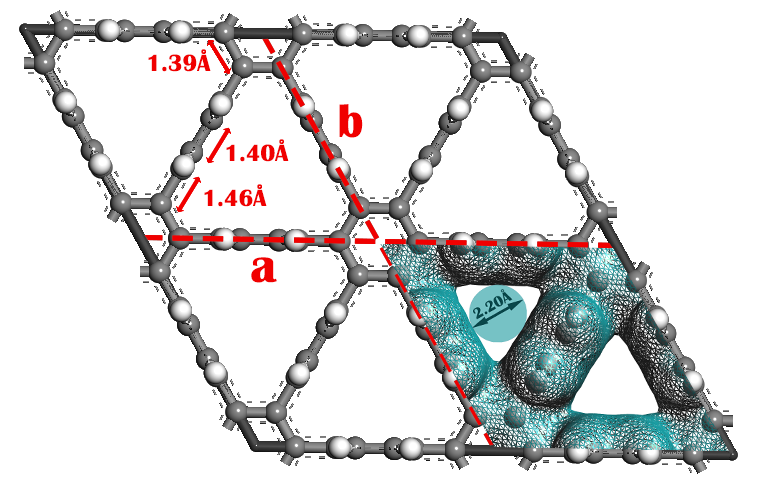}
\caption{The optimized $2 \times 2$ supercell structure of graphenylene--1 monolayer membrane. The electron density
iso--surface of graphenylene--1 monolayer is also shown at isovalue of \SI{0.007} {\elementarycharge\angstrom^{-3}}.}
\label{fcrys}
\end{figure}

The stability of the designed membranes is a crusial factor for their experimental applications. So, we calculate the cohesive energy of graphenylene--1 monolayer membrane to investigate the stability of it. 

The cohesive energy representing the energy required to decompose the membrane into isolated atoms, is defined as \cite{Perim}
\bq
E_{\coh}=\frac{\sum n_{\x} E_{\x}-E_{\T}}{\sum n_{\x}}
\eq
where $n_{\x}$ is the number of atom $\x$ in the membrane, $E_{\x}$ and $E_{\T}$ denote the isolated atom $\x$ and the total energies of the membrane, respectively. Our DFT calculations show that the cohesive energy of graphenylene--1 membrane is \SI{6.52}{\electronvolt}/atom which is a little smaller than that of graphene (\SI{7.95}{\electronvolt}/atom), but is much higher than that of silicene (\SI{3.71}{\electronvolt}/atom) \cite{Li 4}. The value of cohesive energy indicates that graphenylene--1 monolayer is a strongly bonded network and are thus rather stable.

\begin{figure}[t]
\centering
\includegraphics[scale=0.8]{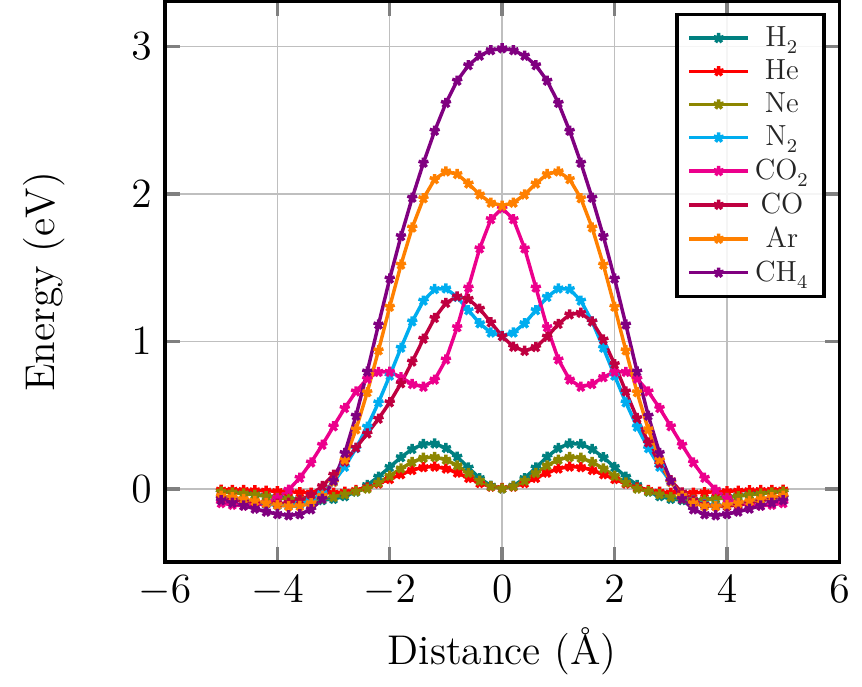}
\caption{Minimum energy pathways for the gases passing through graphenylene--1 membrane in the distance \SI{\pm5}{\angstrom} from the center of the pore.}
\label{comp}
\end{figure}

The interaction energy between the gas molecules and membrane can be obtained as
\bq
E_{\Int}=E_{\gas + \sheet}-(E_{\gas}+E_{\sheet})
\eq
where $E_{\gas + \sheet}$, $E_{\gas}$ and $E_{\sheet}$ are the total energy of the gas molecule adsorbed on graphenylene--1 monolayer, the energy of isolated gas molecule and the energy of graphenylene--1 mononolayer, respectively. In FIG. \ref{comp}, the minimum energy pathways for the gas molecules (including \ce{H2}, \ce{He}, \ce{CO}, \ce{CH4}, \ce{CO2}, \ce{N2}, \ce{Ne} and \ce{Ar}) passing through the membrane are plotted in the distance \SI{\pm5}{\angstrom} from the center of the pore.

\begin{figure*}[t]
\centering
\includegraphics[scale=0.30]{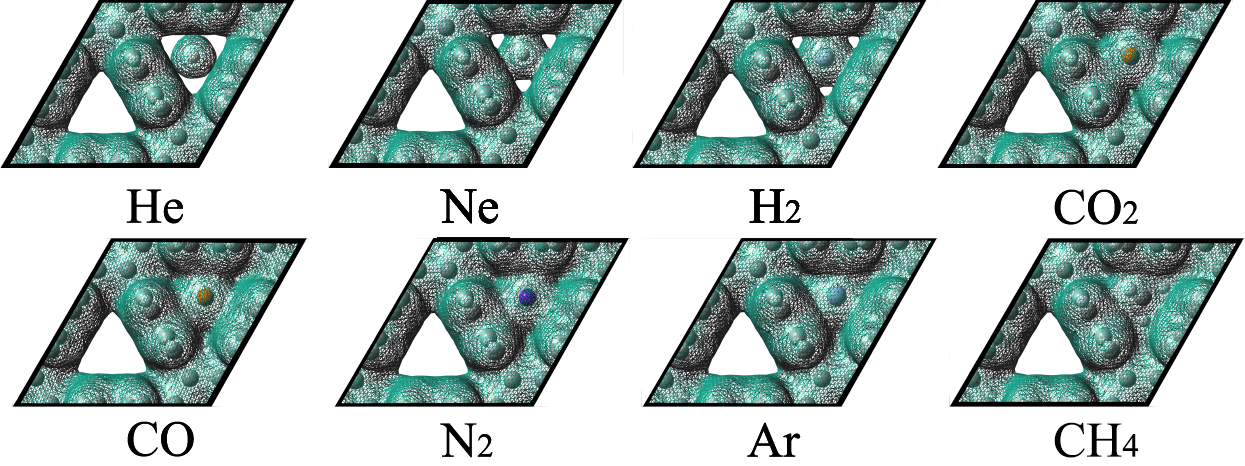}
\caption{Iso--electron density surfaces for the gases passing through graphenylene--1 membrane (isovalue of \SI{0.007}{\elementarycharge\angstrom^{-3}}).}
\label{orb}
\end{figure*}

\begin{table*}[!b]
\small
\centering
  \caption{Comparison of \ce{He} selectivities over impurity gases of graphenylene--1 membrane with other proposed porous membranes at room temperature (\SI{300}{\kelvin})}
  \label{tbl:selHe}
  \begin{tabular*}{\textwidth}{@{\extracolsep{\fill}}lccccccc}
    \hline
membrane & \ce{He}/\ce{Ne} & \ce{He}/\ce{H2} & \ce{He}/\ce{CO2} & \ce{He}/\ce{Ar} & \ce{He}/\ce{N2} & \ce{He}/\ce{CO} & \ce{He}/\ce{CH4} \\
    \hline
{graphenylene--1 (This work)} & \num{1d2} & \num{3d4} & \num{8d30} & \num{1d35} & \num{5d21} & \num{5d21} & \num{2d50} \\
CTF-0 \cite{Wang} & \num{4d6} & \num{4d2} & \num{4d16} & \num{5d35} & \num{2d27} & \num{5d24} & \num{6d38} \\
Silicene \cite{Hu}& \num{2d3} & --- & --- & \num{2d18} & --- & --- & --- \\
Polyphenylene \cite{Blankenburg} & \num{6d2} & \num{9d2} & \num{6d15} & \num{1d30} & \num{2d22} & \num{4d20} & --- \\
Graphdiyne \cite{Bartolomei}& \num{27} & --- & --- & --- & --- & --- & \num{1d24} \\
g-C$_3$N$_4$ \cite{Li 1}& \num{1d10} & \num{1d7} & --- & \num{1d51} & \num{1d34} & \num{1d30} & \num{1d65} \\
C$_2$N$_7$ \cite{Zhu}& \num{3d3} & --- & \num{8d18} & \num{4d18} & \num{3d12} & --- & \num{7d31} \\
    \hline
  \end{tabular*}
\end{table*}

Furthermore, in order to investigate the process which the gases passing through the membrane, we define the diffusion energy barrier for the gas molecules as
\bq
E_{\barrier}=E_{\TS}-E_{\IIS}
\eq
where $E_{\barrier}$, $E_{\TS}$ and $E_{\IIS}$ denote the diffusion energy barrier, the total energy of the gas molecules and the pore center of graphenylene--1 membrane at the transition state and the steady state, respectively. The kinetic diameters ($D_0$) of the studied gases, the adsorption height ($H_{\ad}$) of the steady state of  the gas molecules and the pore center of graphenylene--1 monolayer with corresponding adsorption energy ($E_{\ad}$) and the energy barriers of the gases passing through graphenylene--1 membrane are given in Table \ref{tbl:kindi}.

As is clear in Table \ref{tbl:kindi}, the diffusion energy barrier enhances with increasing the kinetic diameter of the most studied gases. Moreover, there are a surmountable value of diffusion energy barrier for \ce{He} (\SI{0.178}{\electronvolt}) and \ce{H2} (\SI{0.384}{\electronvolt}) molecules passing through the membrane. The calculated adsorption energies of the gases are in the range of \SI{-0.027}{\electronvolt} to \SI{-0.119}{\electronvolt} and the corresponding adsorption heights are in the range of \SI{3.175}{\angstrom} to \SI{4.550}{\angstrom}, which demonstrate the physical adsorption nature of the studied gases on the pore center of graphenylene--1 membrane. The adsorption heights of impurity gases are higher than that of \ce{He} and \ce{H2} molecules. It show that \ce{He} and \ce{H2} molecules are closer to the pore center of the membrane than other gases. The adsorption energy of \ce{He} and \ce{H2} molecules are \SI{-0.027}{\electronvolt} and \SI{-0.077}{\electronvolt}, respectively which are smaller than of other gas molecules except \ce{Ne} molecule. So, \ce{He} and \ce{H2} molecules are easier to desorb from the membrane and pass through it.

\begin{table*}[t]
\small
\centering
  \caption{Comparison of \ce{H2} selectivities over impurity gases of graphenylene--1 membrane with other proposed porous membranes at room temperature (\SI{300}{\kelvin})}
  \label{tbl:selH2}
  \begin{tabular*}{\textwidth}{@{\extracolsep{\fill}}lcccc}
    \hline
    membrane & \ce{H2}/\ce{CO2} & \ce{H2}/\ce{N2} & \ce{H2}/\ce{CO} & \ce{H2}/\ce{CH4} \\
    \hline
    {graphenylene--1 (This work)} & \num{3d27} & \num{2d18} & \num{1d17} & \num{6d46} \\
    {$\gamma$-GYH} \cite{Sang}& \num{9d17} & \num{1d26} & \num{7d23} & \num{2d49} \\
    {Graphenylene} \cite{Song}& \num{1d14} & \num{1d13} & \num{1d12} & \num{1d34} \\
    {$\gamma$-GYN} \cite{Sang}& \num{2d13} & \num{2d21} & \num{1d18} & \num{2d46} \\
    {Porous graphene} \cite{Jiang}& --- & --- & --- & \num{1d22} \\
    {g-C$_2$O} \cite{Zhu 2}& \num{3d3} & \num{2d6} & \num{2d5} & \num{4d23} \\
    {Graphdiyne} \cite{Zhang}& --- & \num{1d3} & \num{1d3} & \num{1d10} \\
    \hline
  \end{tabular*}
\end{table*}

Moreover, we plotted iso--electron density surfaces in FIG. \ref{orb} to study the electron overlaps between the gas molecules and graphenylene--1 monolayer membrane. At least electron overlap between \ce{He} molecule and the membrane causes the energy barrier for \ce{He} among other gas molecules to be the lowest. In addition, more electron overlap between \ce{CH4} molecule and graphenylene--1 membrane makes the highest energy barrier for \ce{CH4} molecule. 

As is well known, the performance of a membrane for the gas separation process is evaluated by the selectivity and the permeance factors. So, we study these factors for the gas molecules passing through graphenylene--1 monolayer sheet.

Regarding the calculated diffusion energy barriers, we can use Arrhenius equation to obtain the selectivity $(S_{\x / \gas})$ for $\x$=He and \ce{H2} over other gas molecules as
\bq
S_{{\x} /{\gas}}=\frac{r_{\x}}{r_{\gas}}=\frac{A_{\x} e^{-E_{\x}/RT}}{A_{\gas}e^{-E_{\gas}/RT}}
\eq
where $r$ is the diffusion rate, $A$ is the diffusion prefactor which is assumed to be identical for all gases ($A$=\SI{1d11}{\per\second}) \cite{Sang}, $E$ is the diffusion energy barrier, $R$ is the molar gas constant and $T$ is the temperature. 

We plotted the calculated selectivities for \ce{He} and \ce{H2} molecules over other gases at different temperatures in FIG. \ref{sel}. As is clear, the selectivity for \ce{He} and \ce{H2} gases decreases with increasing temperature. Moreover, the calculated selectivities for \ce{He} and \ce{H2} molecules over impurity gases for graphenylene--1 membrane and other proposed porous membranes at room temperature (\SI{300}{\kelvin}) are compared in Table \ref{tbl:selHe} and Table \ref{tbl:selH2}, respectively. It can be concluded from Table \ref{tbl:selHe}, graphenylene--1 membrane shows high selectivity for \ce{He} molecule among other proposed membranes especially towrad \ce{H2}, \ce{CO2} and \ce{CH4} gases. Also, the selectivity of \ce{H2}/\ce{CO2}, \ce{H2}/\ce{N2}, \ce{H2}/\ce{CO} and \ce{H2}/\ce{CH4} are \num{3d27}, \num{2d18}, \num{1d17} and \num{6d46}, respectively in Table \ref{tbl:selH2}. The results represent that graphenylene--1 membrane shows high selectivity for \ce{He} and \ce{H2} molecules in comparision with other proposed membranes.

\begin{figure}[!b]
\centering
\subfigure[]{
\includegraphics[scale=0.8]{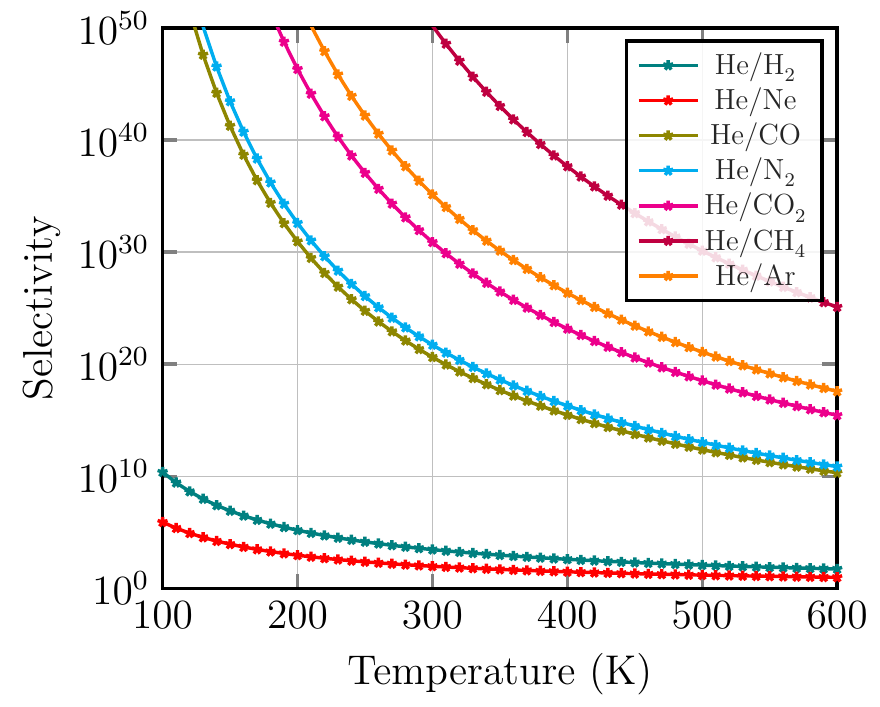}}
\subfigure[]{
\includegraphics[scale=0.8]{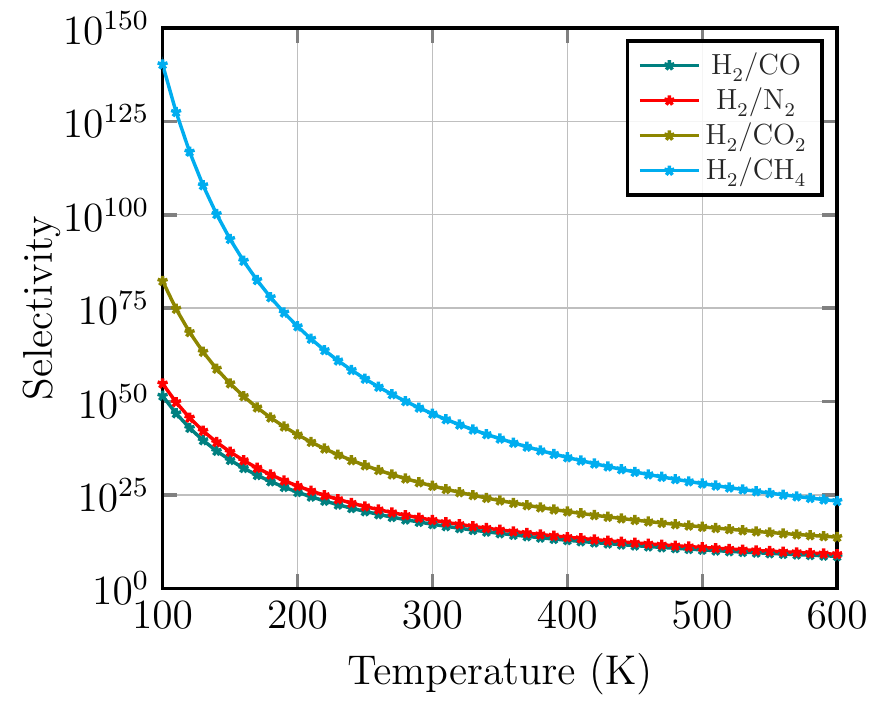}}
\caption{Selectivities of graphenylene--1 membrane for (a) \ce{He} and (b) \ce{H2} molecules over other gases as a function of temperature.}
\label{sel}
\end{figure}

Besides the selectivity, the permeance factor which determines the separation efficiency, is another important factor to characterize the performance of a membrane. 

Considering the calculated energy barriers, we further used the kinetic theory of the gases and the Maxwell-Boltzmann velocity distribution function to investigate the permeances of the gas molecules passing through graphenylene--1 membrane. The number of gas particles colliding with graphenylene--1 monolayer can be obtained as
\bq
N=\frac{P}{\sqrt{2\pi MRT}}
\eq
where $P$ is the gas pressure, here it is taken as \num{3d5} Pa, $M$ is the molar mass, $R$ is the molar gas constant and $T$ is the gas temperature. The probability of a particle diffusing through the pore of the membrane can be expressed as
\bq
f=\int_{v_B}^{\infty} f(v) \D v
\eq

\begin{figure}[t]
\centering
\includegraphics[scale=0.8]{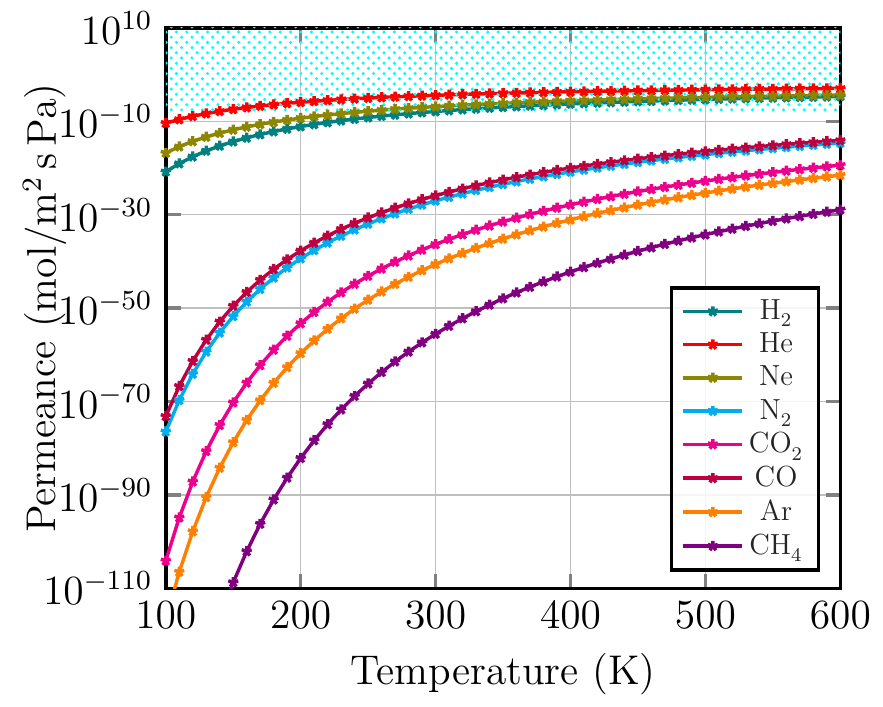}
\caption{Permeances of the studied gases passing through graphenylene--1 membrane as a function of temperature. The cyan dotted area indicates the industrially acceptable standard for the gas separation process.}
\label{per}
\end{figure}

\begin{figure}[t]
\centering
\includegraphics[scale=0.7]{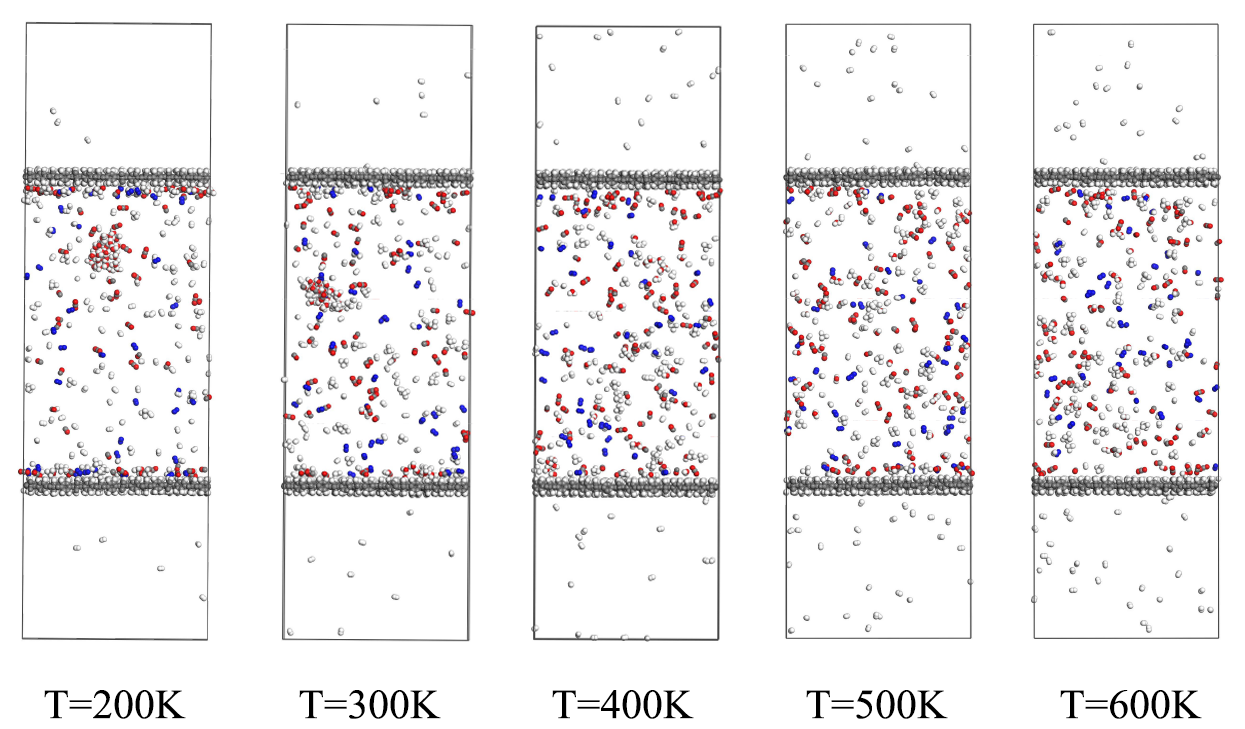}
\caption{Final MD simulated configurations of the mixed gases permeating through the graphenylene membrane at different temperature.}
\label{MD}
\end{figure}

\begin{figure}[!b]
\centering
\includegraphics[scale=0.8]{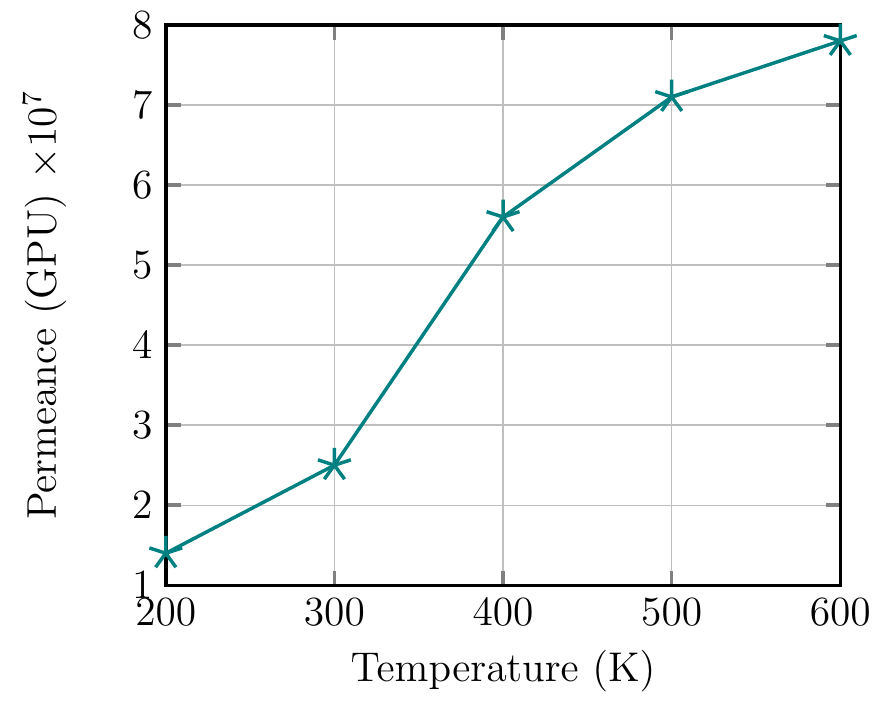}
\caption{Permeance of \ce{H2} molecule passing through graphenylene--1 membrane which obtained from MD simulations, as a function of temperature.}
\label{MDper}
\end{figure}

\begin{figure*}[!b]
\centering
\includegraphics[scale=0.6]{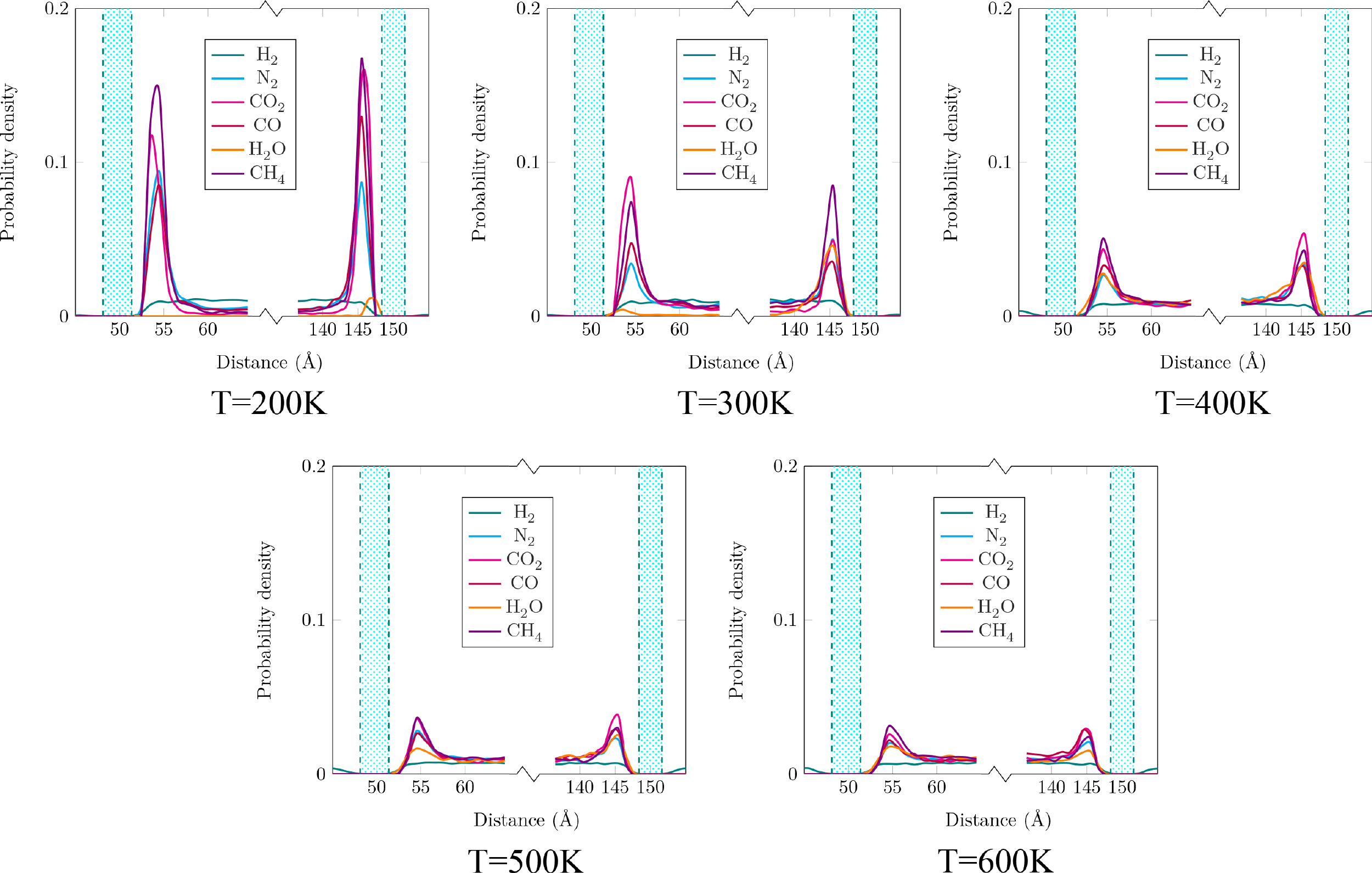}
\caption{Probability density distribution of the gases, as a function of distance to graphenylene--1 monolayer plane. The cyan dotted area represent graphenylene--1 monolayer position.}
\label{ads}
\end{figure*}

where $v_B$ denotes the velocity corresponding to the energy barrier and $f (v)$ is the Maxwell velocity distribution function. So, the flux of the particles can be written as $F=N \times f$ \cite{Ji}. We assume that pressure drop $\Delta P$ equals \num{1d5} \si{\pascal}. Then, we can obtain the permeance factor of the gas molecules passing through the membrane by $p=F/\Delta P$.

In FIG. \ref{per}, we plotted the permeances of the studied gases passing through graphenylene--1 membrane as a function of temperature. The cyan dotted area indicates the industrially acceptable standard for gas separation. As is clear in FIG. \ref{per}, with the increase of temperature, the permeance of each gas increases largely, while the divergence of permeances between different gases decreases. Also, we conclude that graphenylene--1 membrane exhibits the permeance of \ce{H2} and \ce{He} gas molecules are much higher than the value of them in the current industrial applications at temperatures above \SI{300}{\kelvin} and \SI{150}{\kelvin} respectively, while the permeances of \ce{CO2},  \ce{N2}, \ce{CO}, and \ce{CH4} are still much lower than the industrial limit even at \SI{600}{\kelvin}.

To confirm the results of DFT calculations, MD simulations were performed to investigate the process of \ce{H2} molecule passing through graphenylene--1 membrane and estimate \ce{H2} permeance in the temperature range of 200--600 \si{\kelvin}. The final MD simulated configurations of the gas molecules passing through the porous graphenylene--1 membrane at different temperatures are shown in FIG. \ref{MD}. The gases adsorbed on the surface of the membrane by the Van der Waals interaction. Then, they linger on the surface for a few picoseconds before passing through the monolayer membrane, since the gas concentration is different between the gas reservoir (containing \ce{H2O}, \ce{CO2}, \ce{N2}, \ce{CO} and \ce{CH4}) and the vacuum space. After \SI{1}{\nano\second} simulation, we can see that there are 8, 15, 34, 42 and 47 \ce{H2} molecules passing through graphenylene monolayer to the vacuum space at 200, 300, 400, 500 and 600 \si{\kelvin}, respectively, while the other molecules cannot penetrate through this monolayer.

Based on MD simulations, the permeance of \ce{H2} molecules can be defined as \cite{Du}
\bq
p=\frac{\nu}{S\times t \times \Delta P}
\eq
where $\nu$ and $S$ denote the moles of the gas molecules which permeated through the membrane and the area of graphenylene--1 membrane, respectively. Also, $t$ is the time duration of the simulations (\SI{1}{\nano\second}) and the pressure drop ($\Delta P$) is set to 1 \si{\bar} across the pore of the membrane. The permeance of \ce{H2} molecule for graphenylene--1 membrane is plotted in FIG. \ref{MDper}. It can be seen that the permeance of \ce{H2} molecule enhance with increasing temperature. Moreover, the calculated \ce{H2} permeance of graphenylene--1 monolayer together with that of the previously proposed porous membrane at room temperature are summarized in Table \ref{tbl:MDper300}. As is clear, graphenylene--1 membrane shows an appropriate permeance for \ce{H2} gas molecules.

\begin{table}[t]
\large
\centering
  \caption{Comparison of  \ce{H2} permeance of  graphenylene--1 membrane with other proposed porous membranes at room temperature (\SI{300}{\kelvin})}
  \label{tbl:MDper300}
\resizebox{\columnwidth}{!}{%
\begin{tabular}{lcccc}
    \hline
membrane & graphenylene--1 & $\gamma$--GYN \cite{Sang} & $\gamma$--GYH \cite{Sang} & g--C$2$O \cite{Zhu 1}\\

 & (This work)  &  &  & \\
    \hline
Permeance & \num{3d7} & \num{3d7} & \num{1d7} & \num{9d6}\\
    \hline
  \end{tabular}
}
\end{table}

Moreover, the probability density distribution of the gas molecules as a function of distance to graphenylene--1 monolayer plane at different temperature are plotted in FIG. \ref{ads}. The probability density distribution plots show high adsorption for the gas molecules in the range of 3 to 4 \si{\angstrom} from the sheet at low temperatures and these results confirmed the adsorption height which was reached by DFT calculations. At the higher temperatures, according to increasing the kinetic energy for the gas molecules, they overcome to the adsorption energy and easily desorbed from graphenylene--1 monolayer membrane. Therefore, the probability distribution for each molecule decreases. 

\section{Conclusion}

The most proposed porous membranes for \ce{H2} purification and \ce{He} separation encounter the selectivity-permenace trade-off problem. On the other hand, many of them were designed just in the theoretical context. So, the study of new membranes for gas separation process seems very necessary.

In this work, we performed DFT calculations and MD simulations to study graphenylene--1 structure, which synthesized by Liu {\it et al.} \cite{Liu2}, as a monolayer membrane for \ce{H2} purification and \ce{He} separation. First, the stability of the membrane is confirmed by calculating of the cohesive energy. Then, we demonstrated that \ce{H2} and \ce{He} molecules can pass through the membrane easily with a surmountable diffusion energy barrier, \SI{0.384}{\electronvolt} and \SI{0.178}{\electronvolt}, respectively. We showed that the energy barriers of the studied gases (\ce{He}, \ce{H2}, \ce{CO2}, \ce{N2}, \ce{CO}, \ce{Ne}, \ce{Ar} and \ce{CH4}) passing through graphenylene--1 membrane are directly related to their electron density overlap. Moreover, regarding the energy barriers for the gases, we analyzed the performance of the membrane for the gas separation process. Our results show that graphenylene--1 membrane demonstrate high selectivity for \ce{H2} and \ce{He} molecules at room temperature which decrease with raising temperature. Also, the permeance of \ce{H2} and \ce{He} molecules are much higher than the value of them in the current industrial applications at temperatures above \SI{300}{\kelvin} and \SI{150}{\kelvin}, respectively and enhance with raising temperature. In addition, we demonstrated that all the results based on MD simulations are in perfect agreement with DFT calculations. Consequently, graphenylene--1 monolayer membrane can be an excellent candidate for \ce{H2} purification and \ce{He} separation since it shows an appropriate balance between the selectivity and the permeance factors. 

This work provides an interesting approach to introduce a new membrane for separation of the gases. We believed that graphenylene--1 membrane can be a good target for experimental studies, which is very important in industry.

\end{document}